\documentclass{emulateapj}

\usepackage{natbib}

\usepackage{graphicx}
\usepackage[space]{grffile}
\usepackage{latexsym}
\usepackage{amsfonts,amsmath,amssymb}
\usepackage{url}
\usepackage[utf8]{inputenc}
\usepackage{fancyref}
\usepackage{xcolor}
\usepackage{hyperref}
\hypersetup{colorlinks=true,pdfborder={0 0 0},}
\usepackage{textcomp}
\usepackage{longtable}
\usepackage{multirow,booktabs}

\def\beq{\begin{equation}}
\def\eeq{\end{equation}}
\def\ba{\begin{eqnarray}}
\def\ea{\end{eqnarray}}
\def\bal{\begin{align}}
\def\eal{\end{align}}

\shorttitle{Waves in massive stars}
\shortauthors{Fuller et al.}

\begin{document}


\title{The spin rate of pre-collapse stellar cores: \\ wave-driven angular
momentum transport in massive stars}


\author{Jim Fuller\thanks{Email: jfuller@caltech.edu}}
\affil{TAPIR, Walter Burke Institute for Theoretical Physics, Mailcode 350-17, California Institute of Technology, Pasadena, CA 91125, USA, \newline
Kavli Institute for Theoretical Physics, University of California, Santa Barbara, CA 93106, USA}
 
\author{Matteo Cantiello}
\affil{Kavli Institute for Theoretical Physics, University of California, Santa Barbara, CA 93106, USA}
 
\author{Daniel Lecoanet}
\author{Eliot Quataert}
\affil{Astronomy Department and Theoretical Astrophysics Center, University of California at Berkeley, Berkeley, CA 94720-3411, USA}

\begin{abstract}

The core rotation rates of massive stars have a substantial impact on
the nature of core-collapse supernovae and their compact remnants. We
demonstrate that internal gravity waves (IGW), excited via envelope
convection during a red supergiant phase or during vigorous late time
burning phases, can have a significant impact on the rotation rate of
the pre-SN core. In typical ($10 \, M_\odot \lesssim M \lesssim 20 \, M_\odot$) supernova
progenitors, IGW may substantially spin down the core, leading to iron
core rotation periods $P_{\rm min,Fe} \gtrsim 30 \, {\rm s}$. Angular
momentum (AM) conservation during the supernova would entail minimum NS
rotation periods of $P_{\rm min,NS} \gtrsim 3 \, {\rm ms}$. In most
cases, the combined effects of magnetic torques and IGW AM transport
likely lead to substantially longer rotation periods. However,
the stochastic influx of AM delivered by IGW during shell burning phases
inevitably spin up a slowly rotating stellar core, leading to a maximum possible core rotation period.
We estimate maximum iron core rotation periods of $P_{\rm max,Fe} \lesssim 5 \times 10^3 \, {\rm s}$ in
typical core-collapse supernova progenitors, and a corresponding spin
period of $P_{\rm max, NS} \lesssim 500 \, {\rm ms}$ for newborn neutron
stars. This is comparable to the typical birth spin periods of most
radio pulsars. Stochastic spin-up via IGW during shell O/Si burning may
thus determine the initial rotation rate of most neutron stars. For a
given progenitor, this theory predicts a Maxwellian distribution in
pre-collapse core rotation frequency that is uncorrelated with the spin
of the overlying envelope.

\end{abstract}

\bibliographystyle{apj}

\section{Introduction}
\label{intro} 

Rotation is a key player in the drama that unfolds upon the death of a massive star. The angular momentum (AM) contained in the iron core and overlying layers determines the rotation rate at core collapse (CC), which could have a strong impact on the dynamics of CC and the subsequent supernova \citep[see e.g][]{MacFadyen_1999,Woosley_2002,Woosley_2006,Yoon_2006}. Rotation may help determine the nature of the compact remnant, which could range from a slowly rotating neutron star (NS) to a millisecond magnetar or rapidly rotating black hole \citep[see e.g.][]{heger:00,Heger_2005}. The former may evolve into an ordinary pulsar, while the latter two outcomes offer exciting prospects for the production of long gamma-ray bursts (GRB) and superluminous supernova. In each of these phenomena, rotating central engines are suspected to be the primary source of power \citep{1993ApJ...405..273W,Kasen_2010,Metzger_2011}.

Despite rotation being recognized as an important parameter controlling the evolution of massive stars \citep{Maeder_2000}, little is known about the rotation rates of the inner cores of massive stars nearing CC. 
The best observational constraints stem from measurements of the rotation rates of the compact remnants following CC. For instance, a few low-mass black hole X-ray binary systems have been measured to have large spins that can only be accounted for by high spins at birth \citep{Axelsson_2011,Miller_2011,Wong_2012}. However, the rotation rates of young NSs show little evidence for rapid rotation ($P \lesssim 10 \, {\rm ms}$) at birth. The most rapidly rotating young pulsars include PSR J0537-6910 ($P=16\,{\rm ms}$) and the Crab pulsar ($P=33\,{\rm ms}$), whose birth periods have been estimated to be $P_i \lesssim 10\,{\rm ms}$ \citep{marshall:98} and $P_i \sim 19 \, {\rm ms}$ \citep{kaspi:02}, respectively. Many young NSs appear to rotate much more slowly, with typical periods of hundreds of ms \citep{lai:96,gotthelf:13,kramer:03,dincel:15}. In general, pulsar observations seem to indicate a broad range of initial birth periods in the vicinity of tens to hundreds of milliseconds \citep{faucher:06,popov:10,popov:12,gullon:14}. Hence, rapidly rotating young NSs appear to be the exception rather than the rule. 

Theoretical efforts have struggled to produce slow rotation rates. In the absence of strong AM transport mechanisms within the massive star progenitor, NSs would invariably be born rotating near break-up \citep{heger:00}. \citet{Heger_2005} and \citet{Suijs_2008} examined the effect of magnetic torques generated via the Tayler-Spruit (TS) dynamo \citep{spruit:02}, and found typical NS spin periods at birth (assuming AM conservation during CC and the ensuing supernova) of $P \sim 10 \, {\rm ms}$. \citet{wheeler:14} implemented magnetic torques due to MRI and the TS dynamo, and were able to reach iron core rotation rates of $P_{c} \sim 500 \, {\rm s}$, corresponding to NS spin periods of $P \sim 25 \, {\rm ms}$. These efforts are promising, but the operation of both mechanisms within stars has been debated \citep[e.g.][]{Zahn_2007}, and theoretical uncertainties abound.

Recent asteroseismic advances have allowed for the measurement of core rotation rates in low-mass red giant stars (\citealt{beck:12,beck:14,mosser:12,deheuvels:12,deheuvels:14}). In these stars, the core rotates much faster than the surface and one cannot assume nearly rigid rotation as suggested in \citet{Spruit_1998}. However, the cores of low-mass red giants rotate much slower than can be explained via hydrodynamic AM transport mechanisms or magnetic torques via the TS dynamo \citep{cantiello:14}. If similar AM transport mechanisms operate in more massive objects, this suggests that the pre-collapse cores of massive stars may rotate slower than predicted by many previous theoretical investigations.

Internal gravity waves (IGW) constitute a powerful energy and AM transport mechanism in stellar interiors. Several studies (\citealt{kumar:97,zahn:97,Kumar_1999,talon:02,talon:03,talon:05,talon:08,charbonnel:05,denissenkov:08,fullerwave:14}, hereafter F14) have found that convectively generated IGW can redistribute large quantities of AM within low-mass stars. IGW may partially account for the rigid rotation of the Sun's radiative interior and the slow rotation of red giant cores, although magnetic torques are also likely to be important (\citealt{denissenkov:08}, F14). IGW may also be important in massive main sequence stars. \cite{rogers:12} and \cite{rogers:13} have found that convectively generated IGW can alter the spin-rate of the stellar photosphere, while \cite{lee:14} propose that convectively generated IGW in B[e]-type stars may instigate outbursts that expel mass into the decretion disk.

The AM redistribution arising from convectively excited IGW stems primarily from two physical effects. The first, more commonly studied effect, is wave filtering via differential rotation, which stems from an IGW-mean flow interaction. The second, less commonly studied effect, is stochastic variations in the AM flux carried by IGW, which arises because the IGW are excited by the stochastic motions of turbulent convection. We shall see that both mechanisms are important at different phases of the lives of massive stars.

Convectively excited IGW have a particularly strong influence on the evolution of massive stars nearing CC. Indeed, after core carbon exhaustion, waves are the most effective energy transport mechanism within radiative zones, as photons are essentially frozen in and neutrinos freely stream out. In two recent papers, \citet{quataert:12} and \citet{shiode:14} (hereafter QS12 and SQ14) showed that the prodigious power carried by convectively excited waves (on the order of $10^{10} L_\odot$ during Si burning) can sometimes unbind a large amount of mass near the stellar surface, and may substantially alter the pre-collapse stellar structure. IGW are ubiquitous in simulations of late burning stages \citep{Meakin_2006,Meakin_2007,meakinb:07}, although existing simulations have not quantified their long-term impact. 

In this paper, we examine AM transport due to convectively excited IGW within massive stars, focusing primarily on AM transport during late burning stages (He, C, O, and Si burning). We find that IGW are generally capable of redistributing large amounts of AM before CC despite the short stellar evolution time scales. In the limit of very efficient prior core spin-down, we show that a stochastic influx of AM via IGW sets a minimum core rotation rate which is comparable to the broad distribution of low rotation rates ($P \lesssim 500 \,{\rm ms}$) observed for most young NSs. We also examine whether IGW can prevent the core from spinning up as it contracts during stellar evolution. We find that IGW may spin down the core substantially, although spin-down via magnetic torques is likely required to reproduce the observed pulsar population.

Our paper is organized as follows. In Section 2, we describe our massive star models, the generation of IGW during various stages of stellar evolution, and the AM they transport. In Section 3, we consider whether IGW can stochastically spin up a very slowly rotating core, attempting to determine a maximum core rotation period. In Section 4, we investigate whether the IGW can spin down the cores of massive stars, attempting to determine a minimum core rotation period. In Section 5, we conclude with a discussion of our results and their implications for CC, supernovae, and the birth of compact objects.

\section{Convectively Excited IGW}
\label{igw}

IGW are generated by convective zones and propagate into neighboring stably stratified regions, carrying energy and AM. To estimate energy and AM fluxes carried by IGW, we use techniques similar to those of F14, QS12, and SQ14. We begin by constructing a sequence of stellar models using the MESA stellar evolution code \citep{paxton:11,paxton:13}. In what follows, we focus on a $M=12 M_\odot$, $Z=0.02$ model that has been evolved to CC, although we also discuss a more massive ($35 \,M_\odot$) model in Section \ref{spinup}. Details on the models can be found in Appendix \ref{model}. For our purposes, the most important model outputs are the local heat flux, convective mach numbers, and life time of convectively burning zones. As in SQ14, we find these quantities correlate most strongly with the helium core mass. Stellar models of larger zero-age main sequence (ZAMS) mass or with more mixing (due to overshoot or rotation) tend to have a higher He core mass and may exhibit different wave dynamics than our fiducial model. Our main goal here is simply to provide a rough estimate of IGW AM fluxes for a typical low-mass ($M \lesssim 20 M_\odot$) progenitor of a NS. 

\begin{figure}[h!]
\begin{center}
\includegraphics[width=0.98\columnwidth]{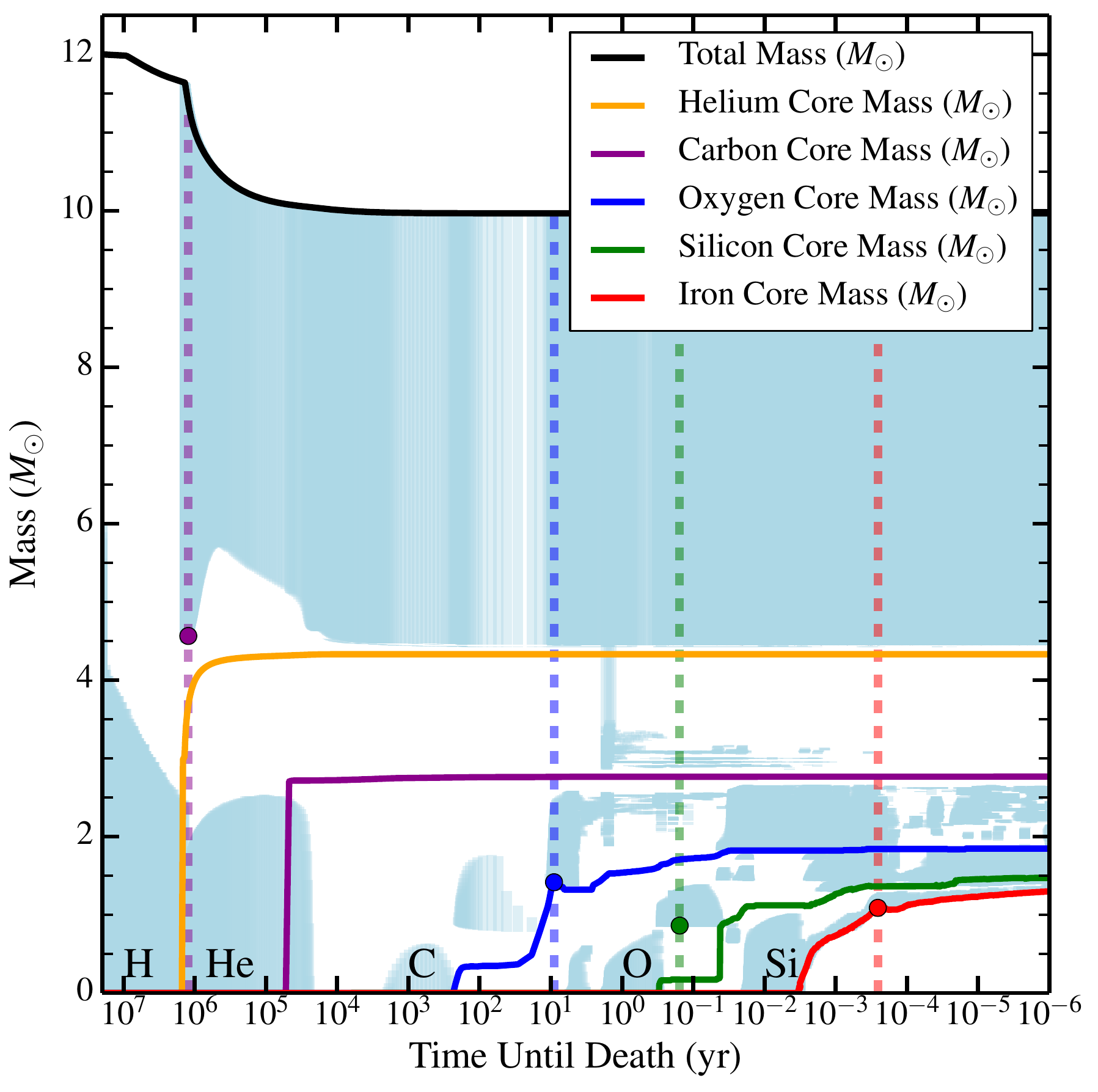}
\caption{\label{fig:MassiveIGWhist} Kippenhahn diagram of our $12 \, M_\odot$ stellar model, with the x-axis showing the time until CC. Light blue shaded regions are convective. Solid colored lines show helium, carbon, oxygen, silicon, and iron core masses. We have labeled convective core burning phases with the element being burned in each phase. Dashed vertical lines show the locations (from left) of our helium core burning, carbon shell burning, oxygen shell burning, and silicon shell burning models used in Figures \ref{fig:Massivestruc} and \ref{fig:MassiveIGWtime}. The mass coordinate of the base of the convective shell, from which IGW propagate downwards, is labeled by a circle on each dashed line.}
\end{center}
\end{figure}

\begin{figure}[h!]
\begin{center}
\includegraphics[width=1\columnwidth]{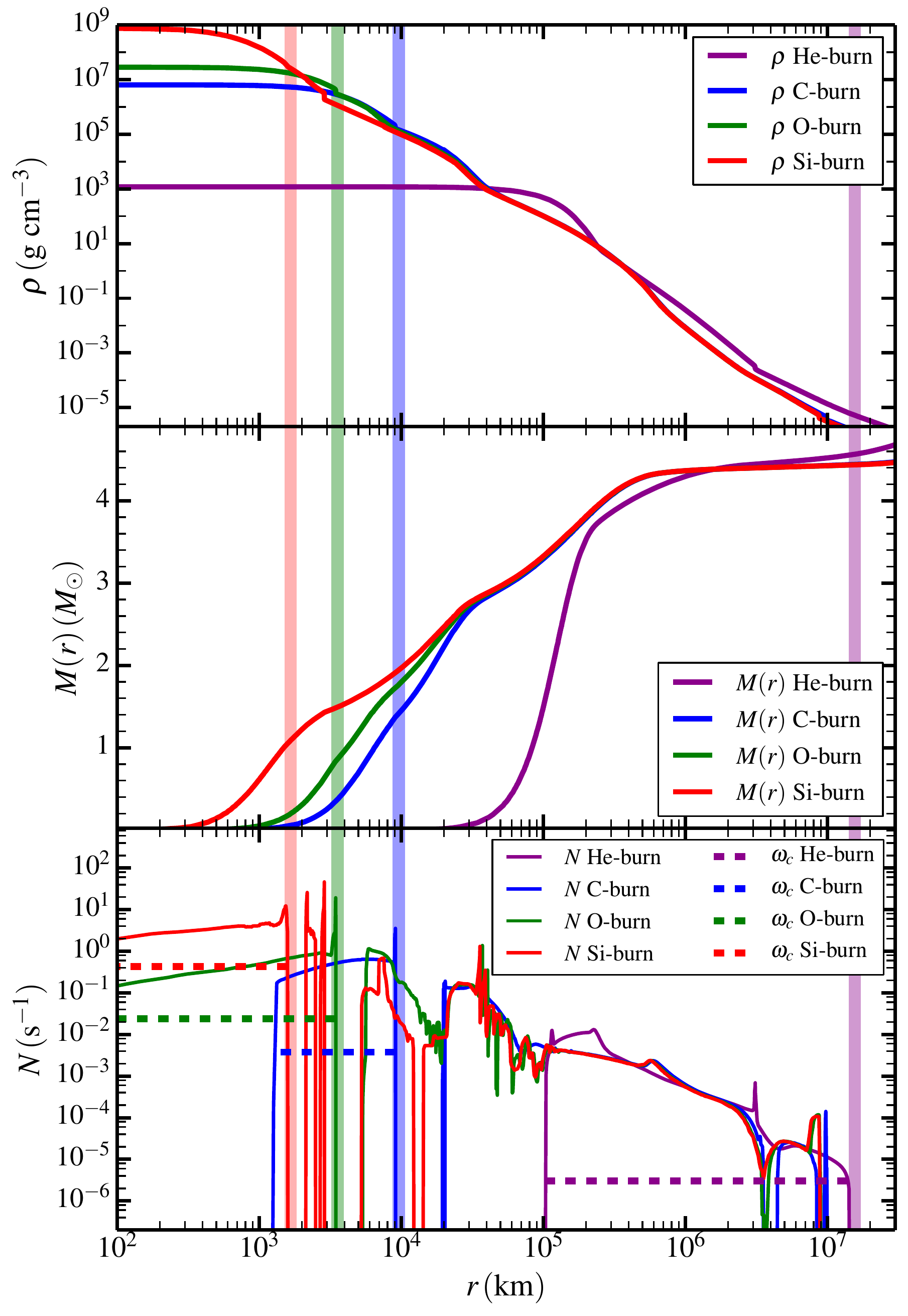}
\caption{ \label{fig:Massivestruc} Internal structure of our $12 M_\odot$ stellar model, at the phases labeled in Figure \ref{fig:MassiveIGWhist}. At all stages shown, the star is a red supergiant with radius $R \sim 10^3 R_\odot$. The He-burn stage is core helium burning, while the C-burn, O-burn, and Si-burn stages are shell carbon, oxygen, and silicon burning phases, respectively. Vertical shaded bars are drawn near the base of the surface convection zone for the He-burn stage, and near the base of the convective shell burning region of the other stages. {\bf Top:} Density profiles $\rho$ of the inner regions of our models. The envelope density profile is similar for each model, but central densities vary by orders of magnitude. {\bf Middle:} Enclosed mass profiles. For C/O/Si burning, the models are chosen such that the mass internal to the convective shell burning region is $M_c \sim 1.2 M_\odot$. {\bf Bottom:} The Brunt-V\"{a}is\"{a}l\"{a} frequency, $N$, of our models. We have also plotted horizontal dashed lines at the convective turnover frequencies $\omega_c$ below the relevant convective zone. IGW propagate in the regions where $\omega_c < N$, shown by the horizontal extent of the dashed lines. }
\end{center}
\end{figure}

A full understanding of AM transport by IGW should include the combined effects of waves emitted from each convective zone. For simplicity, we focus on cases in which a convective shell overlies the radiative core, irradiating it with IGW. These convective shell phases typically occur after core burning phases and thus have the final impact for a given burning phase. We use mixing length theory (MLT), as described in F14, to calculate IGW frequencies and fluxes. Encouragingly, our MLT calculations yield convective velocities and Mach numbers similar to those seen in simulations. We find convective velocities during Si burning very similar to those of \cite{couch:15}. Our $12 M_\odot$ O burning model yields convective velocities and turnover frequencies a factor of $\sim \! 2$ smaller than those seen in simulations of O burning shells (e.g., \citealt{Meakin_2006,meakinb:07,Meakin_2007,Arnett_2008}). However, this is likely due to the larger mass progenitor of their model, as we find higher mass models ($\sim 20 M_\odot$ as used in the works above) yield O burning properties closer to those seen in the simulations. We proceed with our MLT results and expect that realistic wave frequencies may differ from those used here by a factor of $\lesssim \! 2$.

Figure \ref{fig:MassiveIGWhist} shows a Kippenhahn diagram for our stellar model, and Figure \ref{fig:Massivestruc} shows the density ($\rho$), mass [$M(r)$], and Brunt-V\"{a}is\"{a}l\"{a} frequency ($N$) profiles of our model during important convective shell phases. The first convective shell phase occurs during He-core burning, at which point the star has evolved into a red supergiant. At this stage, IGW are generated at the base of the surface convection zone and propagate toward the He burning core. We have also shown profiles during shell C burning, O burning, and Si burning, when the radiative core contains a mass of $M_c \sim 1 M_\odot$ and is being irradiated by IGW generated from the overlying convective burning shell. The basic features of each of these phases is quite similar, the main difference is that more advanced burning stages are more vigorous but shorter in duration. We find that the characteristics of the convective burning shells (convective luminosities, turnover frequencies, mach numbers, and lifetimes) are similar to those listed in QS12 and SQ13, although the shell burning phases are generally more vigorous and shorter-lived than the core burning examined in SQ13. Table 1 lists some of the parameters of our convective zones.

\begin{table*}
\caption{\label{tab:table}}
\begin{center}
\begin{tabular}{ccccccc}
\hline\
Burning Phase & $r_c$ (km) & $T_{\rm shell}$ (s) & $t_{\rm waves}$ (s) & $\mathcal{M}$ & $L_c \, (L_\odot) $ & $\omega_{\rm c}$ (rad/s)  \\
\hline
He Core Burn & $ 1.6 \times 10^7 $ & $ 4 \times 10^{13}$ & $ 2 \times 10^{5}$ & $0.06$ & $6 \times 10^4$ & $ 3 \times 10^{-6}$ \\
\hline
C Shell Burn & $ 9.7 \times 10^3$ & $ 3 \times 10^8$ & $ 10^{6}$ & $0.002$ & $3 \times 10^8$ & $ 4 \times 10^{-3}$ \\
\hline
O Shell Burn & $ 3.6 \times 10^3$ & $ 4 \times 10^6$ & $ 10^{5}$ & $0.004$ & $8 \times 10^9$ & $ 2 \times 10^{-2}$ \\
\hline
Si Shell Burn & $ 1.7 \times 10^3$ & $ 7 \times 10^3$ &  $ 2 \times 10^{3}$ &  $0.02$ & $2 \times 10^{12}$ & $ 4 \times 10^{-1}$ \\
\hline
\end{tabular}
\end{center}
{\small Properties of convection during the late burning phases shown in Figures and \ref{fig:MassiveIGWhist} and \ref{fig:Massivestruc}. Here, $r_c$ is the radius at the base of the convection zone in consideration, $T_{\rm shell}$ is the duration of the burning phase, $t_{\rm waves}$ is a wave spin-up timescale (Equation \ref{eqn:twave}), $\mathcal{M}$ is the convective Mach number, $L_c$ is the luminosity carried by the convective zone, and $\omega_{\rm c}$ is the angular convective turnover frequency.}
\end{table*}

\begin{table*}
\caption{\label{tab:table2}}
\begin{center}
\begin{tabular}{ccccccc}
\hline\
Burning Phase & $r_c$ (km) & $T_{\rm shell}$ (s) & $t_{\rm waves}$ (s) & $\mathcal{M}$ & $L_c \, (L_\odot) $ & $\omega_{\rm c}$ (rad/s)  \\
\hline
C Shell Burn & $ 2.0 \times 10^4$ & $ 7 \times 10^6$ & $ 10^{4}$ & $0.004$ & $9 \times 10^9$ & $ 3 \times 10^{-3}$ \\
\hline
O Shell Burn & $ 4.6 \times 10^3$ & $ 9 \times 10^5$ & $ 5 \times 10^{3}$ & $0.008$ & $2 \times 10^{11}$ & $ 4 \times 10^{-2}$ \\
\hline
Si Shell Burn & $ 2.3 \times 10^3$ & $ 5 \times 10^3$ &  $ 4 \times 10^{2}$ & $0.02$ & $6 \times 10^{12}$ & $ 3 \times 10^{-1}$ \\
\hline
\end{tabular}
\end{center}
{\small Same as Table \ref{tab:table}, but for our $35 \, M_\odot$ model. We do not examine core He burning for this model because it is not a red supergiant and does not have a convective envelope at this stage.}
\end{table*}

The total energy flux carried by waves emitted from the bottom of the convective zone is of order
\begin{equation}
\label{eqn:Ewaves}
\mathcal{M} L_{\rm c} \lesssim \dot{E} \lesssim \mathcal{M}^{5/8} L_{\rm c}  
\end{equation}
\citep{Goldreich_1990,Kumar_1999}, where $\mathcal{M}$ is the convective Mach number (defined as the ratio of MLT convective velocity to sound speed, $v_{\rm c}/c_s$), and $L_{\rm c}$ is the luminosity carried by convection near the base of the convective zone. Many previous works have used the left-hand side of equation \ref{eqn:Ewaves} as an estimate for the IGW energy flux, although \cite{Lecoanet_2013} argue that a more accurate estimate may be $\dot{E} \sim \mathcal{M}^{5/8} L_{\rm c}$, which is larger by a factor of $\mathcal{M}^{-3/8}$. We consider the left-hand side of equation \ref{eqn:Ewaves} to be a lower limit for the wave flux, and the right-hand side to be an upper limit.

For shell burning, this energy flux is dominated by waves with horizontal wave numbers and angular frequencies near $\bar{m} \sim {\rm max}(r_c/H_c,1)$, and $\bar{\omega} \sim \omega_{\rm c}$, respectively. Here, $H_c$ and $r_c$ are the pressure scale height and radial coordinate near the base of the convective zone, and we define the angular convective turnover frequency as
\begin{equation}
\omega_{\rm c} = \frac{ \pi v_{\rm c}}{ \alpha H_c} \, ,
\end{equation}
where $\alpha H_c$ is the mixing length. Our models use $\alpha = 1.5$. For the convective shells we consider, $H_c \sim r_c$ and we expect waves of low angular degree to be most efficiently excited. The characteristic AM flux carried by these waves is 
\begin{equation}
\label{eqn:Jwaves}
\dot{J} \sim \frac{\bar{m}}{\bar{\omega}} \dot{E}.
\end{equation}
Turbulent convection generates waves with a spectrum of azimuthal numbers $m$ and angular frequencies $\omega$. The values given above are characteristic values which dominate the AM flux. The waves carrying the most AM flux sometimes damp before they reach the core, and might not be able to affect the spin of the core. Then the waves with $\bar{m}$ and $\bar{\omega}$ would not dominate the AM flux to the core; instead, other waves in the turbulent spectrum become important (see Section \ref{spindown}). 

As a first check to see if IGW can have any affect on the spin of the core of the star, we assume all waves can propagate to the core. We suppose that IGW could be important for the spin evolution if they are able to carry an amount of AM comparable to that contained in a young NS, which contains $J_{NS} \approx 10^{48} \, g \, {\rm cm}^2 \, {\rm s}^{-1}$ for a rotation period of $P_{\rm NS} = 10 \, {\rm ms}$. Then the characteristic timescale on which waves could affect the AM of the core is
\begin{equation}
\label{eqn:twave}
t_{\rm waves} = \frac{J_{\rm NS}}{\dot{J}} \, .
\end{equation}
Table 1 lists shell burning lifetimes $T_{\rm shell}$ and wave spin-alteration time scales $t_{\rm waves}$, evaluated using $\dot{E} \propto \mathcal{M}$, $\bar{m}=1$ and $\bar{\omega}=\omega_c$. In all phases, $t_{\rm waves} \ll T_{\rm shell}$ for our model, indicating that waves may be able to have a substantial impact on the spin rate of the core. We examine this impact in the following sections.

\section{Stochastic Spin-up by Internal Gravity Waves}
\label{spinup}

The core of a massive star undergoing shell burning is irradiated by IGW, and so it will generally contain a non-zero amount of AM and have a non-zero rotation frequency. Our goal here is to calculate the net AM deposited by IGW in an initially non-rotating core, and therefore to calculate a minimum core spin rate following each shell burning phase.

Our calculation is based on two key ideas. First, we show in Appendix \ref{wavestar} that the core will absorb all of the AM carried inward by IGW because the waves become non-linear due to geometric focusing near the center of the star. This process has been extensively discussed in the literature and shown to occur in hydrodynamical simulations \citep{press:81,barker:10,barker:11}. Section \ref{spindown} also demonstrates that the IGW generated by late shell burning (C burning and beyond) are essentially unattenuated by radiative diffusion or neutrino damping as they propagate inward (see Appendix \ref{tgrow} for additional discussion on the effects of radiative diffusion). IGW emitted from a convective shell will therefore propagate toward the center of the star, non-linearly break, and deposit their AM within the core. The total AM of the core at the end of a shell burning phase will therefore be a superposition of the AM deposited by IGW throughout the shell burning phase.

The second key point is that IGW are generated by the turbulent, stochastic motions of the convective burning shell. The AM flux carried by IGW therefore varies stochastically on a convective turnover time scale. For sufficiently long time scales we expect a nearly zero net AM flux because equal amounts of prograde and retrograde waves will be generated, but on convective turnover time scales the instantaneous AM flux will be non-zero. The net AM flux deposited by IGW during the finite duration of the shell burning phase will therefore also be non-zero. This discreteness of the AM flux is particularly important in late stage shell burning phases because they are short lived compared to main sequence burning phases. We note that stochastic IGW AM transport has also been invoked in \cite{rogers:12} and \cite{lee:14} to modify the AM contained in the atmospheres of stars on the main sequence.

A complication is that IGW absorption at critical layers could occur if IGW generate large shear near the center of the star due to its low moment of inertia. These critical layers can prevent IGW from entering the core if they can migrate outward to the core boundary, which occurs on the time scale $t_{\rm shear} = I_c \omega_c/\dot{J}$. We find $t_{\rm shear} \sim 2 \times 10^{4} \, {\rm s}$  for Si shell burning, using the conservative estimate from equation \ref{eqn:Ewaves}. Although IGW may generate significant shear within the $T_{\rm shell} \sim 7 \times 10^{3} \, {\rm s}$ Si shell burning life time, the shear will not encompass the entire core and prevent further influx of IGWs. We do not attempt to calculate the precise distribution of AM within the core (i.e., the core rotation profile, which will depend on the non-linear wave breaking dynamics and AM redistribution within the core), but only calculate the net AM deposited by IGW within the core.

To quantify our arguments, we consider a non-spinning core being irradiated by IGW emitted from an overlying convective shell. We define an IGW ``packet" as the IGW generated over a convective turnover time $\tau_c = 2 \pi/\omega_c$. Since the convective motions are only correlated over $\sim \! \tau_c$, each wave packet will have an AM vector uncorrelated with the previous wave packet. The spherical symmetry of the background structure implies that these AM vectors will be randomly oriented in space. From equation \ref{eqn:Jwaves}, it follows that each wave packet will have an AM vector of approximate length
\begin{equation}
\label{eqn:AMvec}
J_w \sim \frac{2 \pi}{\bar{\omega}} \dot{J}  \sim \frac{2 \bar{m} \pi}{\omega_{\rm c}^2} \mathcal{M} L_{\rm c} \, ,
\end{equation}
The expected magnitudes of the $x$, $y$, and $z$ components of the AM vector are $J_w/\sqrt{3}$. Since our goal is to find a minimum rotation rate, we use $\bar{m}=1$ to minimize the AM carried by each wave packet.

As IGW packets deposit their AM in the core, the total core AM in each direction exhibits a random walk. During each shell burning phase, the convective shell emits roughly $N$ wave packets, with 
\begin{equation}
\label{eqn:Nwave}
N \sim \frac{\omega_{\rm c} T_{\rm shell}}{2 \pi} \, ,
\end{equation}
where $T_{\rm shell}$ is the length of the shell burning phase. For Si shell burning, $N \sim 400$. This relatively small number demonstrates that the stochastic nature of the wave emission process can be important during late burning phases (in contrast to, e.g., He burning when $N \sim 2 \times  10^7$). After $N$ steps of the random walk, the core AM in each direction has a Gaussian distribution centered around zero and with standard deviation 
\begin{equation}
\label{eqn:Jmean}
\sigma_{J} = \sqrt{ \frac{N}{3} } J_w.
\end{equation}
The magnitude of the total AM vector then has a Maxwellian distribution, with standard deviation $\sigma_J$. The corresponding spin frequency also has a Maxwellian distribution, given by 
\begin{equation}
\label{eqn:jmaxwell}
f(\Omega) = \sqrt{\frac{2}{\pi}} \frac{\Omega^2}{\sigma_\Omega^3} e^{- \Omega^2/(2 \sigma_\Omega^2)} \,.
\end{equation}
where $\sigma_\Omega = \sigma_J/I_c$, and $I_c$ is the moment of inertia of the radiative core. A little algebra yields the corresponding spin period distribution,
\begin{equation}
\label{eqn:pdist}
f(P) = \sqrt{\frac{2}{\pi}} \frac{\sigma_P^3}{P^4} e^{- \sigma_P^2/(2 P^2)} \,,
\end{equation}
with $\sigma_P = 2 \pi/\sigma_\Omega$.

\begin{table*}
\caption{\label{tab:table3}}
\begin{center}
\begin{tabular}{ccccccc}
\hline\
Burning Phase & $P_{\rm ex}$ (s) & $P_{\rm ex,Fe}$ (s) & $P_{\rm ex,NS}$ (s) & $P_{\rm min}$ (s) & $P_{\rm min,Fe}$ (s) & $P_{\rm min,NS}$ (s) \\
\hline
He Core Burn & - & - & -& $ 2 \times 10^{5}$ & $ 30 $ & $ 3 \times 10^{-3} $  \\
\hline
C Shell Burn & $2 \times 10^{4}$ & $4 \times 10^{2}$ & $4 \times 10^{-2}$ & $ 2 \times 10^{3}$ & $ 40$ & $ 4 \times 10^{-3} $  \\
\hline
O Shell Burn & $2 \times 10^{4}$ & $10^{3}$ & $10^{-1}$ & - & - & -  \\
\hline
Si Shell Burn & $4 \times 10^{3}$ & $2 \times 10^{3}$ & $2 \times 10^{-1}$ & - & - & -  \\
\hline
\end{tabular}
\end{center}
{\small Approximate maximum and minimum rotation periods enforced by IGW after the burning phases shown in Figure \ref{tab:table} for our $12 \, M_\odot$ model. The maximum rotation periods $P_{\rm ex}$ are calculated from equation \ref{eqn:pex} using a conservative IGW flux, while the minimum rotation periods $P_{\rm min}$ are calculated in Section \ref{spindown}. $P_{\rm ex}$ is most plausibly set by Si shell burning because magnetic torques cannot compete with IGW during this phase. For each phase of evolution, we list the rotation period immediately after the burning phase, the corresponding rotation rate of the iron core assuming AM conservation (Fe subscript), and the corresponding rotation rate of a NS (NS subscript).}
\end{table*}

\begin{table*}
\caption{\label{tab:table4}}
\begin{center}
\begin{tabular}{cccc}
\hline\
Burning Phase & $P_{\rm ex}$ (s) & $P_{\rm ex,Fe}$ (s) & $P_{\rm ex,NS}$ (s) \\
\hline
C Shell Burn & $3 \times 10^{3}$ & $50$ & $3 \times 10^{-3}$ \\
\hline
O Shell Burn & $2 \times 10^{3}$ & $150$ & $10^{-2}$ \\
\hline
Si Shell Burn & $2 \times 10^{3}$ & $400$ & $3 \times 10^{-2}$ \\
\hline
\end{tabular}
\end{center}
{\small Same as Table \ref{tab:table3}, but listing the values of $P_{\rm ex}$ for our $35 \, M_\odot$ model.}

\end{table*}

The expected angular spin frequency corresponding to equation \ref{eqn:jmaxwell} is
\begin{equation}
\label{eqn:omex}
\Omega_{\rm ex} = \sqrt{\frac{4 \omega_c T_{\rm shell} }{3 \pi^2}} \frac{J_w}{I_c} \, ,
\end{equation}
with a corresponding expected spin period of
\begin{equation}
\label{eqn:pex}
P_{\rm ex} = 8/\Omega_{\rm ex} \, .
\end{equation}
The expected spin frequency scales as $\Omega_{\rm ex} \propto \omega_c^{-1/2} L_c T_{\rm shell}^{1/2}$. Thus, more energetic and long-lived convection yields higher expected rotation rates. The short duration $T_{\rm shell}$ of later burning phases largely counteracts their increased vigor, and we find that earlier burning phases are generally capable of depositing more AM.

\begin{figure*}
\begin{center}
\includegraphics[width=2\columnwidth]{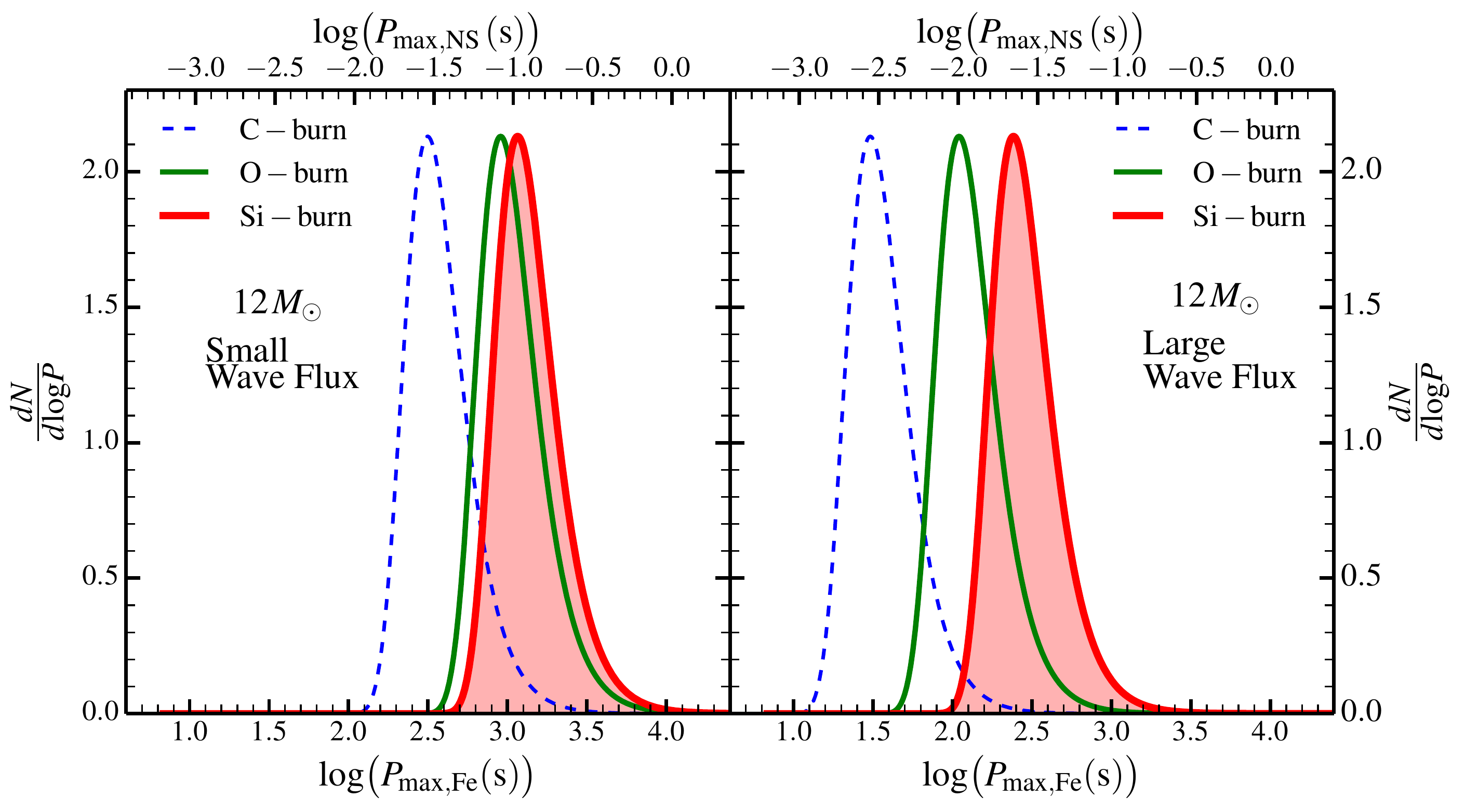}
\caption{ \label{fig:MassiveIGWspin}  Normalized distribution of maximum spin periods in the pre-collapse iron core, $P_{\rm max,Fe}$, due to stochastic spin-up via IGW in our $12 \, M_\odot$ model. The left panel shows the distribution using the pessimistic wave flux (left hand side of equation \ref{eqn:Ewaves}), while the right panel shows the distribution using the optimistic wave flux (right hand side of equation \ref{eqn:Ewaves}). The top axis shows the corresponding maximum NS rotation period, $P_{\rm max,NS}$, assuming conservation of AM during the supernova. $P_{\rm max,NS}$ is most plausibly set during Si burning (see text), which corresponds to the shaded red area. Stochastic IGW spin-up leads to values of $P_{\rm max,NS}$ of tens to hundreds of milliseconds, similar to the initial rotation rates of most pulsars. }
\end{center}
\end{figure*}

We consider the estimate of equation \ref{eqn:omex} to be robust against many uncertainties associated with IGW generation. It is not sensitive to the details of the IGW spectrum, and only relies on the fact that the spectrum is peaked near frequencies of $\omega_c$. Moreover, the scaling above shows that the value of $\Omega_{\rm ex}$ depends weakly upon the convective turnover frequency $\omega_c$. The main uncertainty is the value of the IGW flux, so we will consider both the optimistic and pessimistic limits of equation \ref{eqn:Ewaves}.

The stochastic spin-up process described above will only occur under certain conditions. First, the core and burning shell must be slowly rotating, or else the stochastic spin-up will have a negligible effect. Second, stochastic spin-up can only proceed as long as $\Omega_{\rm ex} \lesssim \omega_c$. If $\Omega_{\rm ex}$ approaches $\omega_c$, wave filtering processes as described in Section \ref{spindown} will alter the subsequent dynamics. Most of our estimates below have $\Omega_{\rm ex} \ll \omega_c$, so we believe they are valid estimates of minimum spin rates. For C burning in the optimistic wave flux estimate ($\dot{E} \propto \mathcal{M}^{5/8}$), however, $\Omega_{\rm ex}$ approaches $\omega_c$, so this value of $\Omega_{\rm ex}$ lies near the maximum rotation rate achievable through stochastic spin-up for our stellar model.

Moreover, stochastic spin-up can only occur if other sources of AM transport (e.g., magnetic torques) operate on longer time scales. This could be the case during late burning phases when magnetic torques become ineffective (\citealt{Heger_2005,wheeler:14}). We can estimate a minimum magnetic coupling time between core and envelope via the Alfven wave crossing time $t_A \approx r_c \sqrt{\rho_c}/B$, with $B$ the approximate magnetic field strength. Typical NS field strengths of a few times $10^{12} \, {\rm G}$ imply field strengths of $\sim 10^8 \,{\rm G}$ in the iron core, which yields $t_A \sim 5 \times 10^4 \,{\rm s}$, much longer than the Si shell burning time (see Table 1). Although magnetic torques may suppress stochastic spin-up during He/C/O burning phases, we expect them to have a negligible impact during Si burning.\footnote{It is possible that the stochastic spin-up process saturates due to back-reaction on the convective shell, which gains AM opposite to the AM deposited by IGW in the core. The induced spin of the convective shell may alter wave generation such that the AM of subsequent wave packets is not randomly oriented and the AM of the core does not undergo a purely random walk. This effect depends on uncertain details such as the rotation profile at the core-shell interface and the effect of the Coriolis force on wave generation. However, we may guess that in the limit of $\Omega_{\rm shell} \ll \omega_c$, the AM of each wave packet could obtain a non-stochastic component of order $ m |J_w| {\bf \Omega}_{\rm shell}/\omega_c$. In this case, the stochastic build-up of AM within the core would saturate when the net deposition of stochastic AM is comparable to the non-stochastic AM deposition, which occurs when $\sqrt{N} \sim N m \Omega_{\rm shell}/\omega_c$. The expected rotation rate of the core after stochastic spin saturation is $\Omega_{\rm ex} \sim \sqrt{2 \pi I_{\rm shell} \dot{E}/\omega_c}/I_{\rm c}$. We find that stochastic spin-up may saturate during C/O burning, leading to $P_{\rm ex,Fe} \sim 2 \times 10^3 \, {\rm s}$ for each of these stages. Interestingly, this spin period is similar to the value of $P_{\rm ex,Fe}$ generated by stochastic spin-up during Si burning. However, the saturation state is unlikely to be reached during Si shell burning, and so our calculations for Si burning remain unchanged. Therefore, we find the value of $P_{\rm ex}$ set by Si burning to be reasonable, regardless of the details of stochastic spin saturation.}

As discussed above, the IGW may generate significant amounts of shear within the iron core, i.e., the core will be differentially rotating at CC. After collapse, we expect the AM to be redistributed within the subsequent NS on a short timescale, e.g., on a NS Alfven time scale $t_A \!\sim \!$ 1 minute. Although differential rotation may persist through CC and the early explosion phase, we expect nearly rigid rotation by the time the NS spin can be observed many years later.

Table \ref{tab:table3} lists maximum expected rotation periods $P_{\rm ex}$ calculated from equation \ref{eqn:pex} after late shell burning phases. We evaluate the core moment of inertia $I_c$ at a mass coordinate of $1.3 \, M_\odot$ for these estimates. We also list the corresponding rotation period $P_{\rm ex,Fe}$ for the pre-collapse iron core, calculated from the value of $I_c$ just before collapse when the mass coordinate at $1.3 M_\odot$ has a radius of $1500 \, {\rm km}$. Additionally, we list the corresponding spin period $P_{\rm ex,NS}$ for a NS with $M_{\rm NS} = 1.3 \, M_\odot$, $R_{\rm NS} = 12 \, {\rm km}$, and $I_{\rm NS} = 0.25 M_{\rm NS} R_{\rm NS}^2$.

Figure \ref{fig:MassiveIGWspin} shows the distribution in maximum spin period of the pre-collapse iron core, $P_{\rm max,Fe}$. We have plotted the values of $P_{\rm max,Fe}$, if the core spin rate is set during C, O, or Si burning.  We find that C, O, and Si burning generate maximum iron core rotation periods in the range of tens to thousands of seconds, depending on the burning phase and efficiency of IGW generation. Si burning most plausibly sets the pre-SN conditions because it is the last convective burning phase before CC and is least likely to be affected by magnetic torques. For the conservative wave flux, we find stochastic spin-up during Si burning leads to $400 \, {\rm s} \lesssim P_{\rm max,Fe} \lesssim  5\times 10^3 \, {\rm s}$. The corresponding NS rotation rate is $40\,{\rm ms} \lesssim P_{\rm NS} \lesssim 500 \, {\rm ms}$. Hence, we find that very slow core rotation rates, as suggested by \cite{Spruit_1998}, are unlikely. Nor do we expect that there is a population of NSs born with very long spin periods, $P \gtrsim 2 \, {\rm s}$, at least from progenitors with ZAMS mass $10 \, M_\odot < M < 20 \, M_\odot$. 

\begin{figure*}
\begin{center}
\includegraphics[width=2\columnwidth]{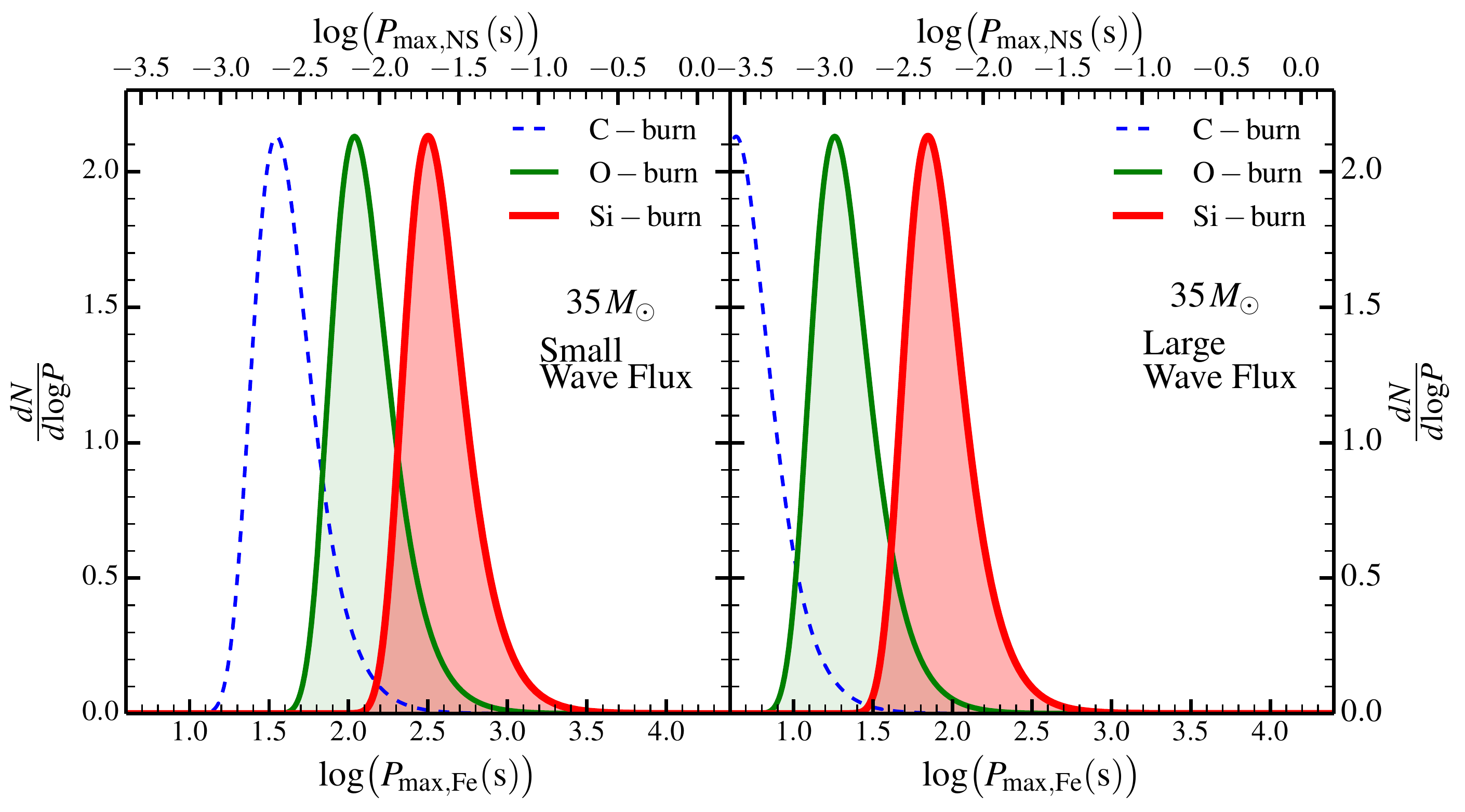}
\caption{ \label{fig:MassiveIGWspin40} Same as Figure \ref{fig:MassiveIGWspin}, but for our $35 \, M_\odot$ progenitor. Larger burning luminosities in this model lead to larger stochastic spin-up rates and smaller values of $P_{\rm max}$. O burning may be more important for this model, but Si burning most likely sets the value of $P_{\rm max}$ (see text). }
\end{center}
\end{figure*}

The distribution of NS spin periods shown in Figure \ref{fig:MassiveIGWspin} appears broadly consistent with those inferred for young NSs \citep{faucher:06,popov:10,gullon:14}. We are therefore tempted to speculate that the stochastic spin-up scenario described above may be the dominant process setting the spin rates of newly born NSs. If so, this scenario predicts that the rotation rate and direction of the NS is uncorrelated with the rotation of the envelope of the progenitor star, in contrast to any sort of magnetic spindown mechanism. However, there are several caveats to keep in mind. First, the scenario presented above can only proceed if the core is initially very slowly rotating, which requires efficient magnetic/IGW core spin-down (see Section \ref{spindown}) to occur before Si burning. Second, the NS rotation rate may be changed during the supernova, by fallback effects, or by the r-mode instability (\citealt{Andersson_1998,Andersson_1999}, see \citealt{ott:2009} for a review). Finally, there is a considerable amount of uncertainty in the IGW energy flux. Since the minimum core rotation rate set by IGW is proportional to the wave energy flux (which is uncertain at an order of magnitude level), there is an equal amount of uncertainty in the induced rotation rates. 
\newline

To understand how stochastic IGW spin-up proceeds in different types of massive stars, we have performed the procedure above for a $35 \, M_\odot$ ZAMS model (described in Appendix \ref{model}), whose late shell burning properties are listed in Table \ref{tab:table2}. The expected spin periods $P_{\rm ex}$ (equation \ref{eqn:pex}) are listed in Table \ref{tab:table4}, and the distribution of expected periods is plotted in Figure \ref{fig:MassiveIGWspin40}. In general, the spin periods of our $35 \, M_\odot$ model are smaller by a factor of a few compared to our $12 \, M_\odot$ model. The reason is that the later burning phases (the O burning phase in particular) are more vigorous (higher $L_c$) and have a shorter duration (lower $T_{\rm shell}$) in higher mass stars with larger He cores. Since $\Omega_{\rm ex} \propto T_{\rm shell}^{1/2} L_c$, the increased convective luminosity wins out, generating higher spin rates in larger mass stars. It is therefore possible that higher mass stars give birth to more rapidly rotating NSs (unless they form black holes instead). For the optimistic wave flux (right panel) or for stochastic spin-up via O shell burning, NS spin periods on the order of milliseconds could be generated via stochastic IGW spin-up. We note that the shorter duration O burning phase of higher mass stars makes it more likely that stochastic spin-up during O shell burning can be preserved until CC. However, as the O shell burning lifetime $T_{\rm shell} \sim 10 \, {\rm days}$ may still be longer than an Alfven crossing time, magnetic torques may suppress stochastic spin-up during C/O burning as discussed above.

Unfortunately, we cannot hope to map out the pre-collapse spin rates of all massive stars in this paper. Different progenitors may yield qualitatively different results. For instance, electron-capture SNe progenitors, whose final stages burning stages are quite different from the CC case, may have different core spin rates. Very massive stars ($M \gtrsim 50 M_\odot$) and stars altered by binary evolution will also make interesting targets for future studies.

\section{Core Spin-down by Internal Gravity Waves}
\label{spindown}

In Section \ref{spinup}, we discussed the spin-up of a non-rotating core during the final shell burning phases before CC. However, stars are born rotating, and in the absence of AM transport the core will spin-up as it contracts during stellar evolution. Our goal here is to determine whether IGW generated during shell burning phases can prevent the spin-up of the core due to contraction. Because we are interested in understanding the effects of IGW on a relatively rapidly rotating stellar core, the random walk arguments of the previous section do not apply.  Instead, as we now explain, the key physics is the preferential absorption of certain waves due to differential rotation-induced wave damping.

Over long time scales, we expect turbulent convection to generate prograde and retrograde waves in nearly equal quantities, so that the net AM flux imparted to IGW is nearly zero. However, differential rotation can set up powerful wave filtration mechanisms (see F14), which filter out either prograde or retrograde waves. Consider waves of angular frequency $\omega$ and azimuthal number $m$ that propagate across a radial region of thickness $\Delta r$, whose endpoints have angular spin frequencies that differ by an amount $\Delta \Omega$. If  $\Delta \Omega > \omega/m$, the waves will encounter a critical layer within the region, and will be absorbed. Only waves of opposing AM will penetrate through the layer; therefore, rapidly rotating regions will only permit influxes of negative AM, which will act to slow the rotation of the underlying layers.

IGW can therefore limit differential rotation to a maximum amplitude $\Delta \Omega_{\rm max} \sim \omega/m$, provided the IGW AM flux is large enough to change the spin rate on time scales shorter than relevant stellar evolution times. In the case of a rapidly rotating core (which has contracted and spun-up) surrounded by a slowly rotating burning shell, we may expect a maximum core rotation rate of $\sim \! \omega$, provided waves of this frequency can propagate into the core. This maximum rotation rate assumes $|m|=1$ waves dominate the AM flux, the actual rotation rate could be smaller if $|m|>1$ waves have a substantial impact. Thus we define a maximum core rotation rate
\begin{equation}
\label{eqn:Omegamax}
\Omega_{\rm max} \sim \omega_* \,,
\end{equation}
where $\omega_*$ is the characteristic frequency of waves that dominate AM transport and are able to penetrate into the core.

The wave frequency $\omega_*$ which dominates AM transport is determined by the IGW frequencies generated by a convective shell, and by the subsequent propagation and dissipation of those waves. In the absence of wave damping, $\omega_* \sim \omega_c$ because these waves dominate the energy/AM flux. However, radiative diffusion preferentially damps low frequency IGW because they have shorter radial wavelengths and slower group velocity. As low frequency IGW damp out, the value of $\omega_*$ shifts to larger frequencies at larger depths below the convective zone. Although radiative dissipation of ingoing IGW is negligible from C shell burning onward, we find it is important for IGW generated by the surface convection zone during the core He burning phase.


To calculate the appropriate value of $\omega_*$, we use the same methods as F14. Upon generation, the IGW carry an energy flux $\dot{E}_0$ and AM flux $\dot{J}_0$, and have a frequency spectrum which is initially peaked around $\omega_c$. We assume the wave spectrum has a power law fall off at higher frequencies such that 
\begin{equation}
\label{eqn:spectrum}
\frac{d \dot{E}_0(\omega)}{d\omega} \sim \frac{\dot{E}_0}{\omega_c} \left(\frac{\omega}{\omega_c}\right)^{-a},
\end{equation}
where $a$ is the slope of the frequency spectrum, which is somewhat uncertain. As in F14, we expect a spectrum slope in the range $3 \lesssim a \lesssim 7$, with a fiducial value of $a=4.5$ as found by \citealt{kumar:99,talon:02}.

F14 show that radiative damping leads to a value of $\omega_*$ of
\begin{equation}
\label{eqn:omstar2}
\omega_*(r) = {\rm max} \bigg[ \omega_c \ , \ \bigg(\frac{4}{a} \int^{r_c}_r dr \frac{\lambda^{3/2} N_T^2 N K }{r^3} \bigg)^{1/4} \bigg] .
\end{equation}
The corresponding AM flux carried by waves of $\omega \! \sim \! \omega_*$ is
\begin{equation}
\label{eqn:Jstar}
\dot{J}_*(r) \sim \bigg[\frac{\omega_*(r)}{\omega_c}\bigg]^{-a} \dot{J}_0.
\end{equation}
In equation \ref{eqn:omstar2}, $r_c$ is the radius of the inner edge of the convective zone, $\lambda = l(l+1)$, $l$ is the angular index of the wave (which corresponds to its spherical harmonic dependence, $Y_{lm}$), $N_T$ is the thermal part of the Brunt-V\"{a}is\"{a}l\"{a} frequency, and $K$ is the thermal diffusivity. In what follows, we focus on $l=1$ waves because they have the longest damping lengths and will dominate the AM flux when the waves are heavily damped. Moreover, focusing on $l=1$ waves allows us to estimate maximum spin frequencies, although slower spin frequencies can be obtained when higher values of $l$ and $m$ contribute to the AM flux.

\begin{figure}[h!]
\begin{center}
\includegraphics[width=1\columnwidth]{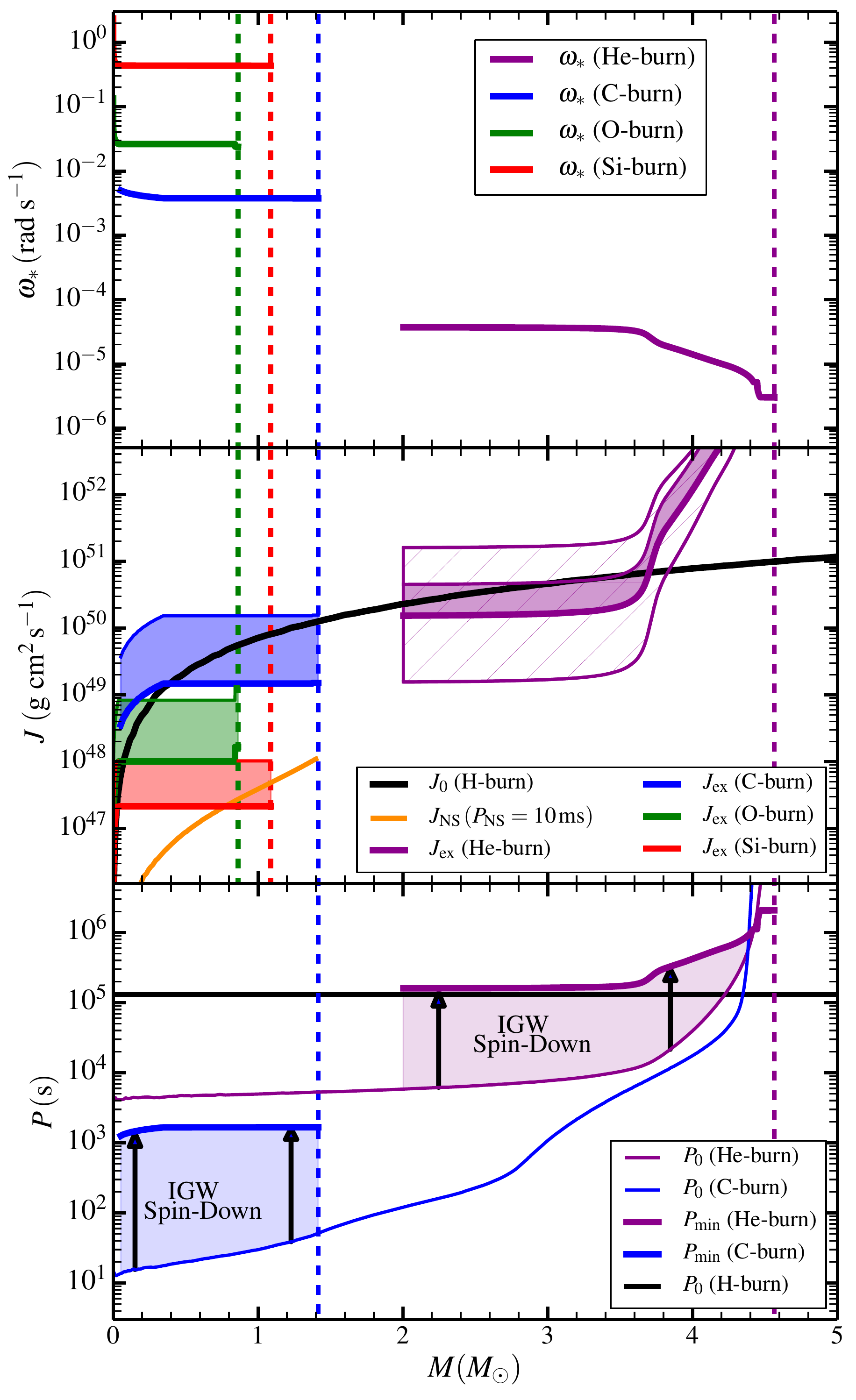}
\caption{ \label{fig:MassiveIGWtime} {\bf Top:} Angular frequency $\omega_*$ of waves that dominate AM transport within our $12 \, M_\odot$ model at different phases of evolution. The dashed vertical lines mark the bottom of the convective zone from which the IGW are launched, while the horizontal extent of each line marks the cavity in which the IGW propagate. We have truncated the curves for He burning at the location of the convective core, into which the IGW cannot propagate. The value of $\omega_*$ increases inward for the He burning model because lower frequency waves are damped by radiative diffusion, although this effect is negligible for C burning and beyond. {\bf Middle:} AM $J_0$ (thick black line) contained within the mass coordinate $M(r)$ of our model rotating with a period of $P_0 \approx 1.5 \, {\rm d}$ on the ZAMS, while the orange line shows the approximate AM $J_{\rm NS}$ contained within $M(r)$ for a NS rotating at $P_{\rm NS} = 10 \, {\rm ms}$. The shaded regions indicate the AM $J_{\rm ex}$ (equation \ref{eqn:Jex}) that can be extracted by IGW during each burning phase. The shaded regions are bounded by lines calculated with a pessimistic and optimistic estimate for IGW fluxes (see equation \ref{eqn:Ewaves}). The hatched region for He burning is calculated using the pessimistic IGW flux for a steep IGW frequency spectrum ($a=5.5$, lower bound) and a shallow spectrum ($a=3.5$, upper bound). IGW can significantly slow the spin rate of the progenitor in regions where $J_{\rm ex} > J_0$.  {\bf Bottom:} Spin periods $P_0$ of our model in the absence of AM transport. As the star evolves, the value of $P_0$ decreases in the contracting core. We have also plotted the approximate minimum spin periods $P_{\rm min}$ which may be enforced by IGW during the core He burning and C-shell burning phases. }
\end{center}
\end{figure}

The top panel of Figure \ref{fig:MassiveIGWtime} shows the value of $\omega_*(r)$ during different burning phases for our $12 \, M_\odot$ model. For C, O, and Si burning, radiative diffusion is negligible and $\omega_* \approx \omega_c$ everywhere. Thus, our calculations are insensitive to uncertainties in the wave spectrum during these stages. $\omega_*$ increases near the center of these models due to non-linear breaking (see Appendix \ref{wavestar}). During He core burning, however, radiative diffusion is very important, causing the value of $\omega_*$ to increase by a factor of $\sim \! 10$ as IGW propagate inward. This decreases the AM carried by the waves and the minimum rotation period they can enforce.


The AM deposited by IGW is only significant if it is larger than the amount of AM contained within the core of the star. A typical massive star has a zero-age main sequence equatorial rotation velocity of $v_{\rm rot} \sim 150 \,{\rm km \ s}^{-1}$ \citep{de_Mink_2013}, corresponding to a rotation period of $P_{\rm MS} \sim 1.5 \, {\rm d}$ for our stellar model. Using this rotation rate, we calculate the AM $J_0(M)$ contained within the mass coordinate $M(r)$, given rigid rotation on the main sequence. In the absence of AM transport, this AM is conserved, causing the core to spin up as it contracts. Of course, magnetic torques may extract much of this AM, so $J_0$ represents an upper limit to the AM contained within the mass coordinate $M(r)$. Both $J_0$ and the corresponding evolving rotation profiles are shown in Figure \ref{fig:MassiveIGWtime}. We also plot the approximate AM $J_{\rm NS}$ contained within a NS rotating at $P_{\rm NS} = 10 \, {\rm ms}$, which is more than two orders of magnitude smaller than the value of $J_0$ within the inner $1.4 M_\odot$. 

The AM capable of being removed from below a radius $r$ via IGW launched during a burning phase is
\begin{equation}
J_{\rm ex} = \eta \dot{J}_*(r) T_{\rm shell} \, .
\label{eqn:Jex}
\end{equation}
Here, $\eta$ is an efficiency factor that accounts for the fact that only some of the wave flux is in low $l$ retrograde waves capable of depositing negative AM in the core. We estimate the value of $\eta$ as follows. For waves with $\omega \sim \omega_c$ launched from a thick convective zone with $r \sim H$, the energy spectrum described in \cite{kumar:99,talon:02,talon:05} has an approximate spectrum $d \dot{E}/dl \propto l e^{-l^2}$. With this spectrum, the energy emitted in $l=1$ waves is approximately one third of the total energy flux. These waves can have azimuthal numbers $m=-1$, $m=0$, or $m=1$, therefore we expect approximately one third of $l=1$ waves are retrograde waves with $m=-1$. Hence, we find $\eta=0.1$ to be a reasonable estimate for the energy flux emitted in low degree retrograde waves capable of propagating into the core. However, we caution that there is a great deal of uncertainty in the wave spectrum, and these numbers should be viewed only as order of magnitude estimates.

Figure \ref{fig:MassiveIGWtime} shows both a pessimistic and optimistic estimate for $J_{\rm ex}$, corresponding to the left and right-hand sides of equation \ref{eqn:Ewaves}, respectively. We find that the values of $J_{\rm ex}$ are comparable to $J_0$ for waves emitted during He core burning and C shell burning. This implies that IGW emitted during these phases may be able to significantly spin down the cores of massive stars. During core He burning, the inner $\sim \! 2 M_\odot$ is convective and IGW cannot propagate into it, hence, it will only be spun down if it is coupled (via its own IGW or via magnetic torques) to the radiative region above it. We have also considered a range of IGW spectra corresponding to $3.5 \leq a \leq 5.5$ in equation \ref{eqn:spectrum}. For steep spectra ($a=5.5$), there is not enough power in high frequency IGW to allow them to spin down the core. For shallow spectra ($a=3.5$), IGW likely can spin down the core. More sophisticated analyses (predicated on a better understanding of the IGW spectrum) are required for a robust conclusion.

During O and Si shell burning, we find that IGW most likely cannot remove the AM contained within the core, if the core retains its full AM from birth. This does not imply IGW have no effect, as the value of $J_{\rm ex}$ for O/Si burning is larger than $J_{\rm NS}$ (the typical AM content of a fairly rapidly rotating NS). Therefore, if the core has been spun down by IGW or magnetic torques during previous burning phases, IGW during late burning phases may be critical in modifying the core spin rate (as discussed in Section \ref{spinup}). 

If IGW are able to spin down the core during He core burning or C shell burning, this entails a minimum possible core rotation period $P_{\rm min} = 2 \pi/\omega_*(r)$ at the end of these phases. The bottom panel of Figure \ref{fig:MassiveIGWtime} plots the value of $P_{\rm min}$, in addition to the rotation profile $P_0$ corresponding to the AM profile $J_0$ that would occur in the absence of AM transport. If IGW are able to spin down the cores, the minimum rotation periods are 10-100 times larger than those that would exist without AM transport. Thus, IGW may significantly spin down the cores of massive stars. We also note that the value of $\omega_*$ is not sensitive the the value of $a$, so $P_{\rm min}$ is insensitive to the IGW spectrum, as long as IGW have enough power to spin down the core. Table \ref{tab:table} lists the values of $P_{\rm min}$ corresponding to He and C burning, as well as corresponding minimium spin periods for the pre-collapse iron core ($P_{\rm min,Fe}$) and for the neutron star remnant ($P_{\rm min,NS}$) given no subsequent AM transport. The minimum NS rotation period $P_{\rm min,NS}$ we calculate is on the order of milliseconds, which is shorter than that inferred for most newly born NSs. Therefore either IGW spin-down is significantly more effective than our estimates, or (perhaps more likely) magnetic torques are responsible for spinning down the cores of massive stars.

\phantom{\,}

\section{Discussion and Conclusions}

We have demonstrated that convectively generated internal gravity waves (IGW) in massive stars are capable of redistributing angular momentum (AM) on short time scales. We have focused primarily on the effects of IGW generated during late stages of massive star evolution (He burning and later) for typical neutron star (NS) progenitors ($10 M_\odot \lesssim M \lesssim 20 M_\odot$). It may seem surprising that AM transport via IGW can act on the short stellar evolution timescales of massive stars nearing core collapse (CC). However, the huge convective luminosities inside evolved massive stars ensure large fluxes of IGW (QS12, SQ14) that can transport energy and AM on short timescales. We therefore encourage efforts to incorporate the effects of IGW in stellar evolution codes focusing on the final stages of massive star evolution.

During He/C burning, inwardly propagating IGW launched from convective shells may slow down the core to much slower spin rates than would be obtained in the absence of other AM transport mechanisms, although the result depends on the IGW spectrum and a robust conclusion remains elusive. If IGW do have an impact, spin-down during core He burning may slow the outer radiative core to minimum spin periods of $P_{\rm min,He} \! \sim \! 2 \, {\rm days}$ in our $12 \, M_\odot$ model. The inner He burning core will also be spun down if it is strongly coupled with the outer core via IGW or magnetic torques. IGW launched during C shell burning may also be able to substantially slow the spin of the core. These convective phases plausibly lead to pre-collapse iron cores with minimum rotation periods $P_{\rm min,Fe} \gtrsim 30\,{\rm s}$, corresponding to initial NS rotation periods of $P_{\rm min,NS} \gtrsim 3 \, {\rm ms}$. The rotation periods listed above are minimum periods for our stellar model. Calculations of rotation rates including magnetic torques \citep{Heger_2005,wheeler:14} typically yield rotation periods several times larger. Magnetic torques may therefore be the dominant AM transport mechanism responsible for extracting AM from massive stellar cores, although it is possible that both mechanisms play a significant role.  

Stochastic influxes of IGW during late burning phases can also lead to the {\em spin-up} of an otherwise very slowly rotating core. This occurs in the case of very efficient prior core spin-down via IGW/magnetic torques. Such efficient core spin-down is not unreasonable, especially given that the cores of low mass red giant stars rotate slower than can be accounted for using existing prescriptions for hydrodynamic mechanisms or magnetic torques via the Tayler-Spruit dynamo \citep{cantiello:14}. It is thus quite plausible that massive star cores are efficiently spun down via IGW/magnetic torques, after which they are stochastically spun up via IGW launched during O/Si burning. This process is similar to previous theories of IGW-induced spin alteration of massive stellar atmospheres \citep{rogers:12}, and stochastic spin-up of proto-NSs during supernovae \citep{spruit:98}. The interesting feature of IGW spin-up during the final stages of massive star evolution is that it may occur on time scales shorter than a core Alfven crossing time and therefore can operate without suppression from magnetic torques.

If this mechanism determines the core spin rate before death, it predicts a Maxwellian distribution in spin frequency, with typical iron core spin periods of $400 \, {\rm s} \lesssim P_{\rm Fe} \lesssim 5 \times 10^3 \, {\rm s}$. We thus find it unlikely that magnetic torques can enforce very large pre-collapse spin periods as claimed by \cite{Spruit_1998}. Additionally, we speculate that the stochastic spin-up process is relatively insensitive to binary interactions or winds that have stripped the stellar envelope, as long as these processes do not strongly modify the core structure and late burning phases. We also express a word of caution, as Si burning is notoriously difficult for stellar evolution codes to handle, and the properties of Si burning produced by our MESA evolutions have large associated uncertainties. The rough energy and AM fluxes in convectively excited waves are, however, reasonable at the order of magnitude level.

If AM is conserved during the supernova, stochastic IGW spin-up entails NS birth periods of $40 \, {\rm ms} \lesssim P_{\rm NS} \lesssim 500 \, {\rm ms}$, albeit with significant uncertainty. These estimates are comparable to initial spin periods of some young NSs (\citealt{lai:96,gotthelf:13}), and to the broad inferred birth spin period distribution of $P_{\rm NS} \lesssim 500 \, {\rm ms}$ for ordinary pulsars (\citealt{faucher:06,popov:10,gullon:14}). Therefore, stochastic IGW spin-up could be the dominant mechanism in determining the rotation periods of pre-collapse SN cores and newborn NSs. In this scenario, there is little or no correlation between the spin of the progenitor and the spin of the NS it spawns. Although torques during the supernova may modify the spin rate of the NS, they would have to be very finely tuned to erase the stochastic spin-up occurring during shell burning. Any sort of purely frictional spin-down processes would likely slow the NS to rotation periods larger than typically inferred for young NSs.

We have also investigated stochastic IGW core spin-up for a $35\,M_\odot$ model. We find that the increased late time burning luminosities of more massive stars leads to more effective stochastic spin-up, creating spin rates a factor of a few higher than those listed above. For optimistic wave fluxes, the initial rotation rates of NSs born from massive progenitors may be as short as several milliseconds. Even larger rotation rates could be possible if magnetic torques are weak enough to allow the stochastic core spin-up generated during O shell burning to persist until CC.

We have quantitatively considered the implications of IGW AM transport in two distinct limits. 
First, we neglect AM transport by magnetic fields and consider the limit in
which the core is rotating much faster than the convective shell because of AM conservation during core contraction. In this case, IGW emitted from
convective shells propagate into the radiative core and may be able to substantially
slow its rotation, enforcing a maximum rotation rate. The second limit we consider is when the stellar core has been
efficiently spun-down via magnetic coupling to the envelope and/or IGW in earlier
phases of stellar evolution. In this case, we have shown that the stochastic influx of AM
via IGW from shell burning leads to a {\em spin-up} of the stellar core and a
minimum core rotation rate. Taken together, these limits enforce iron core rotation periods $30 \, {\rm s} \lesssim P_{\rm Fe} \lesssim 5 \times 10^3 \, {\rm s}$ and initial NS rotation periods of $3 \, {\rm ms} \lesssim P_{\rm NS} \lesssim 500 \, {\rm ms}$. We expect these limits to be robust against many uncertain factors in massive star evolution, e.g., birth spin rate, mass loss, mixing, and the effects of magnetic fields.

There is ample evidence that {\it some} CC events occur with rapidly rotating cores. In particular, long GRBs almost certainly require a rapidly rotating central engine \citep{1993ApJ...405..273W,Yoon_2006,Woosley_2006,Metzger_2011}, and the picture advanced above must break down in certain (although somewhat rare) circumstances. It is not immediately clear what factors contribute to the high spin rate in GRB progenitors, as our analysis was mostly restricted to ``typical" effectively single NS progenitors with $10 M_\odot \lesssim M \lesssim 20 M_\odot$, which explode to produce type-IIp supernovae during a red supergiant phase \citep[see e.g.][]{Smartt_2009}. We speculate that GRB progenitors (if occurring in effectively single star systems) have {\it never} undergone a red supergiant phase, as torques via magnetic fields and/or IGW are likely to spin down the helium core by coupling it with the huge AM reservoir contained in the slowly rotating convective envelope. Our preliminary examination of more massive models indicates that stochastic spin-up may lead to larger spin rates, but is unlikely to generate very rapid rotation throughout the entire He core. A third possibility is that spin-up via mass transfer/tidal torques in binary systems is required for GRB production \citep{Cantiello_2007}. A merger or common envelope event after the main sequence could also remove the extended convective envelope and prevent it from spinning down the core.

The population of massive stars approaching death is complex, and factors such as initial mass, rotation, metallicity, binarity, magnetic fields, overshoot, mixing, winds, etc., will all contribute to the anatomy of aging massive stars. We have argued that AM transport via convectively driven IGW is likely to be an important factor in most massive stars. But it is not immediately obvious how this picture will change in different scenarios, e.g., electron capture supernovae, very massive $(M \gtrsim 50 M_\odot)$ stars, interacting binaries, etc. We hope to explore these issues in subsequent works.

\section{Acknowledgments}

This paper was written collaboratively, on the web, using \href{https://www.authorea.com}{Authorea}. We thank Dave Arnett, Sean Couch, and Christian Ott for useful discussions. JF acknowledges partial support from NSF under grant no. AST-1205732 and through a Lee DuBridge Fellowship at Caltech. DL is supported by a Hertz Foundation Fellowship and the National Science Foundation Graduate Research Fellowship under Grant No. DGE 1106400. EQ was supported in part by a Simons Investigator award from the Simons Foundation and the David and Lucile Packard Foundation. This research was supported by the National Science Foundation under grant No. NSF PHY11- 25915 and by NASA under TCAN grant No. NNX14AB53G.

\begin{appendix}

\section{Stellar Models}
\label{model}

Our stellar models are constructed using the MESA stellar evolution code \citep{paxton:11,paxton:13}, version 6794. The models are run using the following inlist controls file.

\begin{verbatim}
\& star_job  
      kappa_file_prefix = 'gs98'  
/ ! end of star_job namelist

\& controls          
     initial_mass = 12.
     initial_z = 0.020           
     sig_min_factor_for_high_Tcenter = 0.01
     Tcenter_min_for_sig_min_factor_full_on = 3.2d9
     Tcenter_max_for_sig_min_factor_full_off = 2.8d9
     logT_max_for_standard_mesh_delta_coeff = 9.0 
     logT_min_for_highT_mesh_delta_coeff = 10  
     delta_Ye_highT_limit = 1d-3
     okay_to_reduce_gradT_excess = .true. 
     allow_thermohaline_mixing = .true.
     thermo_haline_coeff = 2.0
     overshoot_f_above_nonburn = 0.035      
     overshoot_f_below_nonburn = 0.01
     overshoot_f_above_burn_h = 0.035
     overshoot_f_below_burn_h = 0.0035
     overshoot_f_above_burn_he = 0.035
     overshoot_f_below_burn_he = 0.0035
     overshoot_f_above_burn_z = 0.035
     overshoot_f_below_burn_z = 0.0035
     RGB_wind_scheme = 'Dutch'
     AGB_wind_scheme = 'Dutch'
     RGB_to_AGB_wind_switch = 1d-4
     Dutch_wind_eta = 0.8
     include_dmu_dt_in_eps_grav = .true.
     use_Type2_opacities = .true.
     newton_itermin = 2
     mixing_length_alpha = 1.5
     MLT_option = 'Henyey'
     allow_semiconvective_mixing = .true.
     alpha_semiconvection = 0.01
     mesh_delta_coeff = 1.
     varcontrol_target = 5d-4
     max_allowed_nz = 10000
     mesh_dlog_pp_dlogP_extra = 0.4
     mesh_dlog_cno_dlogP_extra = 0.4      
     mesh_dlog_burn_n_dlogP_extra = 0.4
     mesh_dlog_3alf_dlogP_extra = 0.4
     mesh_dlog_burn_c_dlogP_extra = 0.10
   	 mesh_dlog_cc_dlogP_extra = 0.10
     mesh_dlog_co_dlogP_extra = 0.10
     mesh_dlog_oo_dlogP_extra = 0.10
     velocity_logT_lower_bound=9
     dX_nuc_drop_limit=5d-3
     dX_nuc_drop_limit_at_high_T = 5d-3 ! 
     screening_mode = 'extended'   
     max_iter_for_resid_tol1 = 3
     tol_residual_norm1 = 1d-5
     tol_max_residual1 = 1d-2  
     max_iter_for_resid_tol2 = 12
     tol_residual_norm2 = 1d99
     tol_max_residual2 = 1d99
     min_timestep_limit = 1d-12 ! (seconds)
     delta_lgL_He_limit = 0.1 !
     dX_nuc_drop_max_A_limit = 52
     dX_nuc_drop_min_X_limit = 1d-4
     dX_nuc_drop_hard_limit = 1d99 
     delta_lgTeff_limit = 0.5  
     delta_lgL_limit = 0.5
     delta_lgRho_cntr_limit = 0.02        
     T_mix_limit = 0
/ ! end of controls namelist

\end{verbatim}

The most important feature of this model is that it contains significant convective overshoot, especially above convective zones. It is non-rotating, thus there is no rotational mixing.

Just before core O burning, we change to a 201-isotope reaction network:
\begin{verbatim}
     change_net = .true.   
     new_net_name = 'mesa_201.net'
\end{verbatim}
Although our choices affect details of the model (e.g., He core mass), the general features of our model are robust. It always explodes as a red supergiant. It always undergoes convective core C burning, followed by shell C burning, core O/Ne burning, shell O burning, core Si burning, shell Si burning, and then CC. The approximate convective properties (as described by MLT) are not strongly affected by model parameters. Since these properties are most important for IGW AM transport, we argue that the general features of IGWs described in this work are fairly robust against uncertain parameters in our massive star models.

Differences in the inlist for our $35 \, M_\odot$ model are 

\begin{verbatim}

&star_job

      relax_initial_Z = .true. ! gradually change abundances, reconverging at each step.
      new_Z = 2d-3

/ ! end of star_job namelist

&controls
      initial_mass = 35.

      okay_to_reduce_gradT_excess = .true. !MLT++

      overshoot_f_above_nonburn = 0.015       
      overshoot_f_below_nonburn = 0.005

      overshoot_f_above_burn_h = 0.015
      overshoot_f_below_burn_h = 0.005

      overshoot_f_above_burn_he = 0.015
      overshoot_f_below_burn_he = 0.005

      overshoot_f_above_burn_z = 0.015
      overshoot_f_below_burn_z = 0.005


      varcontrol_target = 8d-4

/ ! end of controls namelist

\end{verbatim}

Notable features of this model are its lower metallicity and the use of MLT++ \cite{paxton:13} to evolve the star through super-Eddington phases of evolution.

\section{Non-linear Damping}
\label{wavestar}

IGW will overturn and break, leading to local energy/AM deposition, if they obtain sufficiently non-linear amplitudes. Here we estimate those amplitudes and the AM flux that can be carried toward the center of the star as waves are non-linearly attenuated. For traveling waves in the WKB limit, it is well known that the radial wave number is
\begin{equation}
\label{eqn:kr}
k_r = \frac{\sqrt{\lambda} N}{r \omega} \, .
\end{equation}
It is straightforward to show that the radial displacement $\xi_r$ associated with IGW of frequency $\omega$ carrying an energy flux $\dot{E}$ is
\begin{equation}
\label{eqn:eflux}
| \xi_r | = \bigg[ \frac{\sqrt{\lambda} \dot{E} }{\rho N r^3 \omega^2} \bigg]^{1/2} \, .
\end{equation}
The waves become non-linear and break when
\begin{equation}
\label{eqn:nl}
| k_r \xi_r | = \bigg[ \frac{\lambda^{3/2} N \dot{E} }{\rho r^5 \omega^4} \bigg]^{1/2} \sim 1 \, .
\end{equation}
In the absence of damping, $\dot{E}$ is a conserved quantity. Therefore, waves become more non-linear as they propagate into regions with larger $N$, lower density, or smaller radius. In our problem, the geometrical focusing (i.e., the $r$-depdendence) is the most important feature of equation \ref{eqn:nl}, and causes waves to non-linearly break as they propagate inward. Note also the $\omega^{-2}$ dependence of equation \ref{eqn:nl}, which causes low frequency waves to preferentially damp.

Equation \ref{eqn:nl} entails there is a maximum energy flux that can be carried by waves of frequency $\omega$, 
\begin{equation}
\label{eqn:emax}
\dot{E}_{\rm max} = \frac{ A^2 \rho r^5 \omega^4}{\lambda^{3/2} N} \, ,
\end{equation}
for waves that non-linearly break when $|k_r \xi_r| = A \sim 1$. We find that IGW of frequency $\omega \sim \omega_c$ become strongly non-linear as they propagate inward during C/O/Si shell burning. Therefore, most of the IGW energy/AM will be deposited within the core. Where the waves are highly non-linear, the waves which dominate the energy flux are those which are on the verge on breaking. To determine their frequency, we use the frequency spectrum of equation \ref{eqn:spectrum} to find
\begin{equation}
\label{eqn:efreq}
\dot{E}_0 \bigg( \frac{\omega}{\omega_c} \bigg)^{1-a} \sim \dot{E}_{\rm max}.
\end{equation}
Solving equation \ref{eqn:efreq} yields the wave frequency which dominates energy transport,
\begin{equation}
\label{eqn:omstarnl}
\omega_* \sim {\rm max} \bigg[ \omega_c \ , \ \omega_c \bigg( \frac{ A^2 \rho r^5 \omega_c^4}{\lambda^{3/2} N \dot{E}_0} \bigg)^{-1/(a+3)}  \bigg] \, .
\end{equation}
We expect frequency spectra with slopes somewhere near $3 \lesssim a \lesssim 7$. Therefore the exponent in equation \ref{eqn:omstarnl} is quite small, and in most cases, $\omega_*$ does not increase to values much larger than $\omega_c$. 

Substituting equation \ref{eqn:omstarnl} back into equation \ref{eqn:emax} allows us to solve for the energy and AM flux as a function of radius due to non-linear attenuation. The result is
\begin{equation}
\label{eqn:jstarnl}
\dot{J}_* \sim \bigg[ \frac{\omega_*(r)}{\omega_c} \bigg]^{-a} \dot{J}_0 \sim {\rm min}\bigg[ \dot{J}_0 \ , \ \bigg(\frac{A^2 \rho r^5 \omega_c^4}{\lambda^{3/2} N \dot{E}_0} \bigg)^{a/(a+3)} \dot{J}_0  \bigg] \, .
\end{equation}

During C/O shell burning, we find that radiative diffusion only slightly damps waves near the shell burning convective zone, while non-linear breaking damps waves near the center of the star. In this case, we first calculate $\omega_*$ and its corresponding energy flux $\dot{E}_*$ via equations  \ref{eqn:spectrum} and \ref{eqn:omstar2}. We then substitute the value of $\dot{E}_*$ for $\dot{E}_0$ in equation \ref{eqn:omstarnl}. The appropriate value of $\omega_*$ is then $\omega_* = {\rm max} \big[ {\rm Eqn.}$ \ref{eqn:omstar2}$, {\rm Eqn.}$\ref{eqn:omstarnl}$ \big]$. The corresponding AM flux is $\dot{J}_* \sim \Big[ \frac{\omega_*(r)}{\omega_c} \Big]^{-a} \dot{J}_0$.

The cores of massive stars nearing death cool primarily through neutrino emission, so it is not unreasonable to think that waves may be damped via neutrino emission. We calculate neutrino energy loss rates in the same manner as \cite{murphy:04}. We find that neutrino damping time scales are always longer than the wave crossing timescale
\begin{equation}
\label{eqn:tcross}
t_{\rm cross} = \int^{r_c}_0 \frac{dr}{v_g}\, ,
\end{equation}
where the IGW radial group velocity is $v_g = r \omega^2/(\sqrt{\lambda} N)$. This is not surprising, as \cite{murphy:04} found neutrino growth/damping rates were slower than stellar evolution time scales. Neutrino damping can therefore be safely ignored.

\section{Wave-Flow Interaction}
\label{tgrow}

It is well known that the interaction between flows (e.g., differential rotation) and IGW can strongly affect IGW propagation and dissipation. Here we show that these interactions will not prevent the stochastic spin-up process (Section \ref{spinup}) from occurring. 

One reason that wave-flow interactions will not prevent stochastic spin-up is that the waves are very weakly damped by radiative diffusion during late burning stages. As shown in Section \ref{spindown}, radiative damping times are long. Nonetheless, the background flows are unstable in the presence of IGW in the sense that IGW will amplify very small amounts of shear. From equation 13 of F14, the timescale on which waves amplify shear due to radiative diffusion is
\beq
t_{\rm grow} \sim \frac{\rho r^4 \omega_* L_d}{\dot{J}_*} \, .
\eeq
where $L_d$ is the wave damping length due to radiative diffusion. In the Sun, this timescale is quite short. However, during late burning stages in the cores of massive stars, it typically exceeds the remaining lifetime of the star. We plot $t_{\rm grow}$ below in Figure \ref{fig:MassiveIGWtgrow}. During O/Si burning, $t_{\rm grow}$ is much longer than the life-time of the star. During C-burning, it is comparable to the remaining lifetime, and therefore IGW may create large amounts of shear during C-burning. 

\begin{figure*}
\begin{center}
\includegraphics[width=0.6\columnwidth]{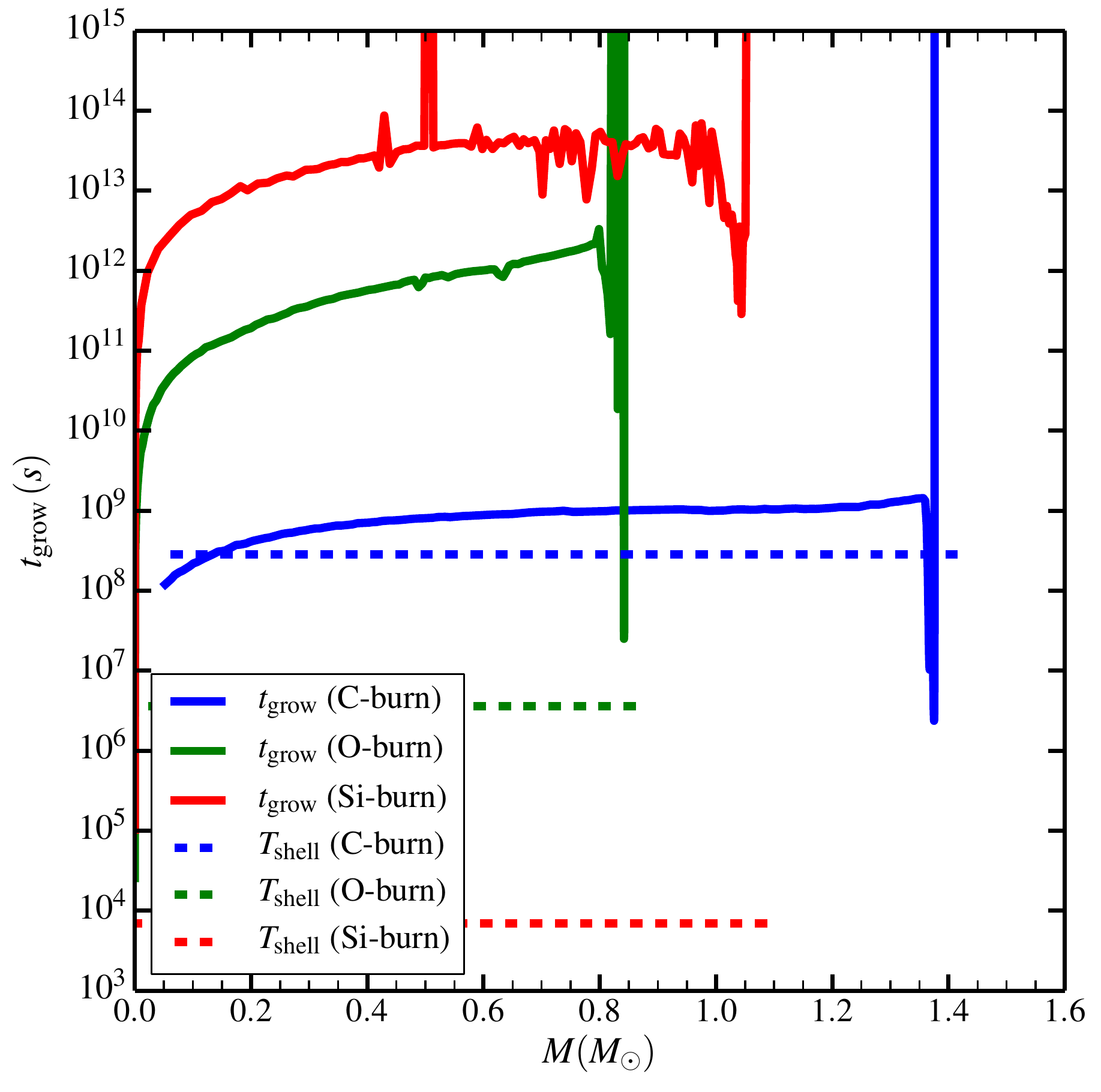}
\caption{ \label{fig:MassiveIGWtgrow}  Growth time for the development of shear, $t_{\rm grow}$, due to radiative diffusion in the radiative cores of our $12 \, M_\odot$ model. For O/Si burning, $t_{\rm grow} \gg T_{\rm shell}$, and it is unlikely that radiative diffusion will cause the IGW to generate significant amounts of shear. }
\end{center}
\end{figure*}

Because $t_{\rm grow}$ is long, IGW will not be able to amplify small amounts of shear into critical layers that absorb subsequent waves and prevent IGW propagation into the core. Therefore, we find that the stochastic spin-up process outlined in Section \ref{spinup} will not be prevented by IGW-flow interaction.

\end{appendix}

\bibliography{converted_to_latex}

\end{document}